\documentstyle[emulateapj,graphicx]{article}
\def\etal  {{et~al.}\ }

\def\msun{{\rm\,M_\odot}}

\def\zsun{{\rm\,Z_\odot}}

\def\vol#1  {{{#1}{\rm,}\ }}

\def\etal{et al.\ }

\def\gsim{\;\rlap{\lower 2.5pt
 \hbox{$\sim$}}\raise 1.5pt\hbox{$>$}\;}
\def\lsim{\;\rlap{\lower 2.5pt
   \hbox{$\sim$}}\raise 1.5pt\hbox{$<$}\;}

\newcount\refno
\newcount\rfno
\def\eq{$^{\the\refno\ }$\advance\refno by 1}
\def\ad{\advance\rfno by 1}

\def\clock{\count0=\time \divide\count0 by 60
     \count1=\count0 \multiply\count1 by -60 \advance\count1 by \time
     \number\count0:\ifnum\count1<10{0\number\count1}\else\number\count1\fi}

\def\myputfigure#1#2#3#4#5%
{\hskip0.03\textwidth\vskip#5pt
\makebox[0pt]{\hskip#2in
\includegraphics[width=#3\textwidth]{#1}}\vskip#4pt\hfill}
\newcommand{\beq}{\begin{equation}}
\newcommand{\eeq}{\end{equation}}

\begin{document}

\title{Synchronized Formation of Sub-Galactic Systems at Cosmological Reionization: Origin of Halo Globular Clusters}
\author{Renyue Cen\altaffilmark{1}}
\altaffiltext{1} {Princeton University Observatory, Princeton University, Princeton, NJ 08544; cen@astro.princeton.edu}

\begin{abstract}

Gas rich sub-galactic halos with mass $M_t\le 10^{7.5}\msun$,
while incapable of forming stars due to lack of adequate coolants, 
contain a large fraction of 
baryonic mass at cosmological reionization.
We show that the reionization of the universe at $z=10-20$ 
has an interesting physical effect on these halos.
The external radiation field causes a synchronous inward 
propagation of an ionization front towards each halo,
resulting in an inward, convergent shock.
The resident gas of mass $M_b\sim 10^4-10^{7}\msun$ 
in low spin (initial dimensionless spin parameter $\lambda\le 0.01$)
halos with a velocity dispersion $\sigma_v\le 11~$km/s
would be compressed by a factor of $\sim 100$ in radius and form
self-gravitating baryonic systems. 
Under the assumption that 
such compressed gaseous systems 
fragment to form stars,
the final stellar systems will have a size $\sim 2-40~$pc,
velocity dispersion $\sim 1-10$km/s and a
total stellar mass of $M_*\sim 10^3-10^{6}\msun$.

The characteristics of these proposed systems seem to
match the observed properties of halo globular clusters.
The expected number density 
is consistent with the observed number density of halo globular clusters.
The observed mass function of slope $\sim -2$ at the high mass end
is predicted by the model.
Strong correlation between velocity dispersion and luminosity 
(or surface brightness) and lack of correlation between velocity
dispersion and size,
in agreement with observations, are expected.
Metallicity is, on average, expected to be low
and should not correlate with any other quantities of globular clusters,
except that a larger dispersion of metallicity among globular clusters
is expected for larger galaxies.
The observed trend of specific frequency with galaxy type
may be produced in the model.
We suggest that these stellar systems
are seen as halo globular clusters today.

\end{abstract}


\section{Introduction}

Zinn (1985) pointed out insightfully that there appears to exist two 
separate populations of globular clusters in the Galaxy:
one which is spherically distributed,
has low metallicity with no radial metallicity gradient
and low rotation, the other which 
is more concentrated towards the Galactic center and 
has high metallicity, a gradient in metallicity and high rotation.
Such a dichotomy seems to be shared by other galaxies
(e.g., 
Ashman \& Bird 1993;
Forbes, Grillmair, \& Smith 1997;
Forbes, Brodie, \& Huchra 1997;
Kundu 1999;
Gebhardt \& Kissler-Patig 1999;
Barmby \etal 2000;
Beasley \etal 2000).
The first population, the halo globular clusters
in both the Galaxy and external galaxies,
display an additional array of interesting properties (Harris 1991).
First, they are old with age $\ge 10$Gyr.
Second, their characteristics are uniform across
all galaxy types and sizes. 
Third, their mass function resembles a power-law 
($\propto M^{-1.7 {\rm\ to\ } -2.0}$) at the high
mass end. 

It has been a long standing challenge to explain 
such 
features.
Several important theoretical models
including the pre-galactic Jeans mass based model (Peebles \& Dicke 1968)
and the secondary thermal instability based model (Fall \& Rees 1985)
have been put forth but 
may not be without some intriguing difficulties.
For example, the Jeans mass based model 
is not naturally expected to produce power-law mass function 
for globular clusters, while the 
thermal instability based model appears to require delicate heating sources
(Harris \& Pudritz 1994; Meylan \& Heggie 1997).
A fully consistent model for the origin 
of the halo globular clusters remains unavailable at this time.
In this paper,
we suggest that the reionization of the universe 
could trigger a simultaneous formation of sub-galactic systems 
at that epoch,
whose characteristics match remarkably well those of
the observed halo globular clusters.

After describing the globular cluster formation theory
in \S 2,
we discuss some of the expected properties of globular clusters
which form in the current theory in \S 3,
followed by conclusions in \S 4.
The following cosmological model is used throughout:
$\Omega_{CDM}=0.26$, $\Omega_{b}=0.04$,
$\Omega_{t}=\Omega_{CDM}+\Omega_{b}=0.3$,
$\Lambda=0.0$, 
$H_0=65$km/sec/Mpc and $\sigma_8=1.0$;
the global gas to total mass ratio $R\equiv \Omega_b/\Omega_t=0.13$.

\section{Formation of Globular Clusters at Cosmological Reionization}

In the standard picture of structure formation,
the universe is thought to be reionized mostly by photons
from quasars or stellar systems more massive 
than $10^9\msun$ (Haiman, Rees \& Loeb 1997;
Gnedin \& Ostriker 1997).
Less massive systems, after having produced a trace amount 
of stars (Pop III) at an earlier epoch,
can no longer form stars
due to lack 
of cooling processes (Haiman, Thoul, \& Loeb 1996;
Haiman, Rees \& Loeb 1997;
Tegmark \etal 1997).
During the cosmological reionization phase
small halos with neutral 
gas sitting idly within (i.e., without internal radiating 
sources) suddenly see a sea of radiation approaching.
We will show that this radiation field has a dramatic dynamic
effect on them.
While an exact treatment of this process is quite intricate and
would require detailed three-dimensional radiation-hydrodynamical 
simulations,
we present a simple analysis that we believe captures
the essence of the process.
To make the calculations relatively tractable,
we assume, without loss of primary characteristic features,
that the halos are initially 
spherical with the following density profile:
\begin{eqnarray}
\rho (r) &=\rho_v ({r\over r_v})^{-2} \quad\quad\quad \hbox{for} \quad\quad\quad r<r_v \nonumber\\
         &=\rho_v ({r\over r_v})^{-3} \quad\quad\quad \hbox{for} \quad\quad\quad r>r_v,
\end{eqnarray}
\noindent 
where $\rho_v\equiv 100\rho_{c}(z)$ is the density at the virial radius
$r_v$ and $\rho_{c}(z)=\rho_c(0)\Omega_t(1+z)^3$ 
is the critical density at redshift $z$
[$\rho_c(0)$ is the critical density today].
The virial mass of the halo (i.e., the total
mass interior to $r_v$) is 
\begin{eqnarray}
M_v = 4\pi \rho_v r_v^3
\end{eqnarray}
\noindent 
and the one-dimensional 
velocity dispersion within the virialized region ($r<r_v$) is
\begin{eqnarray}
\sigma_v = \sqrt{2\pi G\rho_v r_v^2},
\end{eqnarray}
\noindent 
where $G$ is the gravitational constant. 
We assume that the ratio of gas to total mass
in the halos under consideration is equal to the global baryon to
total mass ratio $R$.
The total mass within the radial range $r_v$ to $r>r_v$ (outside the virial
radius) is
\begin{eqnarray}
M(r_v\rightarrow r) = M_v \ln {r\over r_v}.
\end{eqnarray}
\noindent 

The clumpy universe is reionized outside in:
low density regions are ionized first and higher density
regions (except those in the most massive, ionizing sources)
become ionized at progressively later times
(Miralda-Escud\'e, Haehnelt, \& Rees 2000; Gnedin 2000b).
The reionization phase progresses on a time scale of
a Hubble time at the redshift in question;
the mean (volume weighted) radiation field builds up
slowly up to a value of approximately
$10^{-24}~$erg/cm$^2$/hz/sec/sr at Lyman limit.
An accelerated phase follows, when about
$90\%$ of the baryons have been ionized and a sudden jump 
in the amplitude of the mean radiation field intensity at Lyman limit
to $10^{-22}-10^{-21}~$erg/cm$^2$/hz/sec/sr 
occurs within a redshift interval of a fraction of unity and
the reionization process is said to be complete
(Miralda-Escud\'e \etal 2000; Gnedin 2000b).
The mean free path of ionizing photons 
approaches the horizon size at a much later time.
As Figure 3 (below) will show that the gas mass contained
in the halos of interest here (including gas surrounding the halos
that will be compressed) is of order $50\%$ of total gas,
so we are concerned with the phase when 
a large fraction of the gas is not yet ionized and 
the radiation field
is still increasing very slowly (see Figure 2a of Gnedin 2000b).

Let us now examine the {\it global} ionization front
(I-front), which propagates
from low density regions towards halos.
The speed of the external global I-front
towards a halo is 
\begin{eqnarray}
v_{I} = {m_p \over f_H R} J_\nu \rho_v^{-1} \left({r\over r_v}\right)^3
\end{eqnarray}
\noindent 
at $r>r_v$, where $J_\nu\equiv \int_{\nu_0}^\infty 4\pi j_\nu d\nu/h\nu $ 
is the number of ionizing photons per cm$^2$~sec
and $j_\nu$ is the intensity 
of the meta-galactic radiation field;
where $f_H=0.76$ is the hydrogen mass fraction of baryons;
$m_p$ is the proton mass.
Adopting a powerlaw
$j_\nu = j_{LL,21} 10^{-21} (\nu/\nu_0)^{-\beta}~$erg/cm$^2$/hz/sec/sr
one obtains 
$J_\nu=1.9\times 10^6\beta^{-1}j_{LL,21}~$cm$^{-2}$~sec$^{-1}$.
Using $\beta=2$ and $R=0.13$ we obtain
\begin{eqnarray}
v_{I} = 1.8 \left({r\over r_v}\right)^3\left({1+z_{ri}\over 16}\right)^{-3}
\left({j_{LL,21}\over 2\times 10^{-3}}\right)~\hbox{km/s},
\end{eqnarray}
\noindent 
where $z_{ri}$ is the redshift when this occurs.
Note that for a softer stellar spectrum of $j_\nu$,
$\beta\sim 4$ and $v_I$ would be lower. 

The I-front is accompanied by a shock front.
In order for the I-front to
drive a strong compressive shock inward,
the shock velocity has to exceed the velocity 
of the ionization front (D-type),
which decreases with decreasing halo-centric radius.
Otherwise, the neutral gas gets ionized and raised to a high pressure
before the gas is reached
by the shock front (see Spitzer 1978 for an introduction on
the subject and Bertoldi \& McKee 1990 for
a detailed treatment in the context of interstellar clouds).
We assume that 
this condition sets the initial radius of the inward shock, $r_i$:
at $r_i$ the ionization front and the shock have the same velocity.
In the present case the shock velocity 
will instantaneously settle to a speed of 
$v_{shk}=4/3 \sqrt{k T_{ri}/m_p}=12(T_{ri}/10^4K)^{1/2}~$km/s
at the start of the shock due to the ram pressure (see equation 7 below),
where $T_{ri}$ is the temperature of photo-ionized gas 
and $k$ is the Boltzmann constant.
We will return to this when we discuss 
gas cooling in the shell later in this section.
Since $v_{shk}\ge 15~$km/s, this initial radius 
is outside the virial radius (see equation 6)
(note that, for a more realistic
density profile where the density slope near the viral radius
is about $-2.4$ instead of $-3$ assumed here, the initial shock radius
would be still larger).

We make some simplifying assumptions to treat such a shock.
We assume that the shock is strong, which is justified
since neutral gas has a very low temperature,
and the material that
is swept up along its way is accumulated in a thin shell 
obeying the momentum equation 
\begin{eqnarray}
{d(M_s v_s)\over dt} = {GM(<r) M_s\over r^2} + 4\pi r^2 (p_{ext}-p_{int}),  
\end{eqnarray}
\noindent 
where 
$M_s$ is the gas mass within the shell;
$v_s$ is the velocity of the shell;
$M(<r)$ is the total mass interior to radius $r$;
the first term and the two terms within the parentheses on the right hand side 
are due to gravity, external and internal pressure, respectively.
In addition,
we assume that dark matter does not respond to the motion of the gas,
which is unlikely to be in serious error for our purpose.
In any case, our assumption is conservative in the sense
that inclusion of dark matter response to the gas collapse
would enhance the collapse of halo gas.

It is instructive to qualitatively 
consider the calculation in two separate regions:
$r_v\le r<r_i$ and $0\le r<r_v$.
At $r_v<r<r_i$ we assume that the gas is initially cold 
(i.e., $p_{int}$ may be ignored) and 
$p_{ext}=R \rho_v (r_i/r_v)^{-3} k T_{ri}/m_p$ is constant.
The shell will be subject to external pressure and
gravity, counter-balanced by ram pressure,
and will obtain a certain velocity when reaching $r=r_v$.
At $0<r<r_v$, if the shell travels supersonically, the downstream pressure
is undisturbed.
Since initially the gas in the virialized region at $0<r<r_v$
is in hydrostatic equilibrium, 
the region interior to the shell at any moment will remain in 
hydrostatic equilibrium.
In other words, for a shell traveling supersonically,
the gravitational force and 
interior pressure force just cancel out.
In the absence of 
the external pressure 
and the self-gravity of the shell 
(which is small until the shell becomes self-gravitating),
the shell will cruise inward (as long as it remains supersonic),
except that it slows down due to the swept-up gas. 

Quantitatively,
we integrate the momentum equation (7) numerically 
from $r_i$ to $r=0$,
with the initial condition $v_s(r_i)=0$ and $M_s=0$.
We define $\eta$ in the following equation and 
require $\eta$ to be greater than $1$ (momentum condition)
in order for the shell to reach the center of the halo:
\begin{equation}
\eta\equiv {\min [v_s(r)]\over\sigma_v} \ge 1,
\end{equation}
\noindent 
where ${\min [v_s(r)]}$ is the minimum velocity of the shell
in the radial range $[0, r_v]$,
and $\sigma_v$, the velocity dispersion, is used as
the isothermal sound speed of the gas at $r<r_v$.
Note that if equation (8) is not met and the shell slows down to
a subsonic speed at some radius (likely at a large radius since
most of the mass is at large radii),
the interior gas will adjust its pressure 
causing the interior gas to be adiabatically compressed,
which will quickly build up the internal pressure and cause 
the shell to slow down and eventually to stop.

\myputfigure{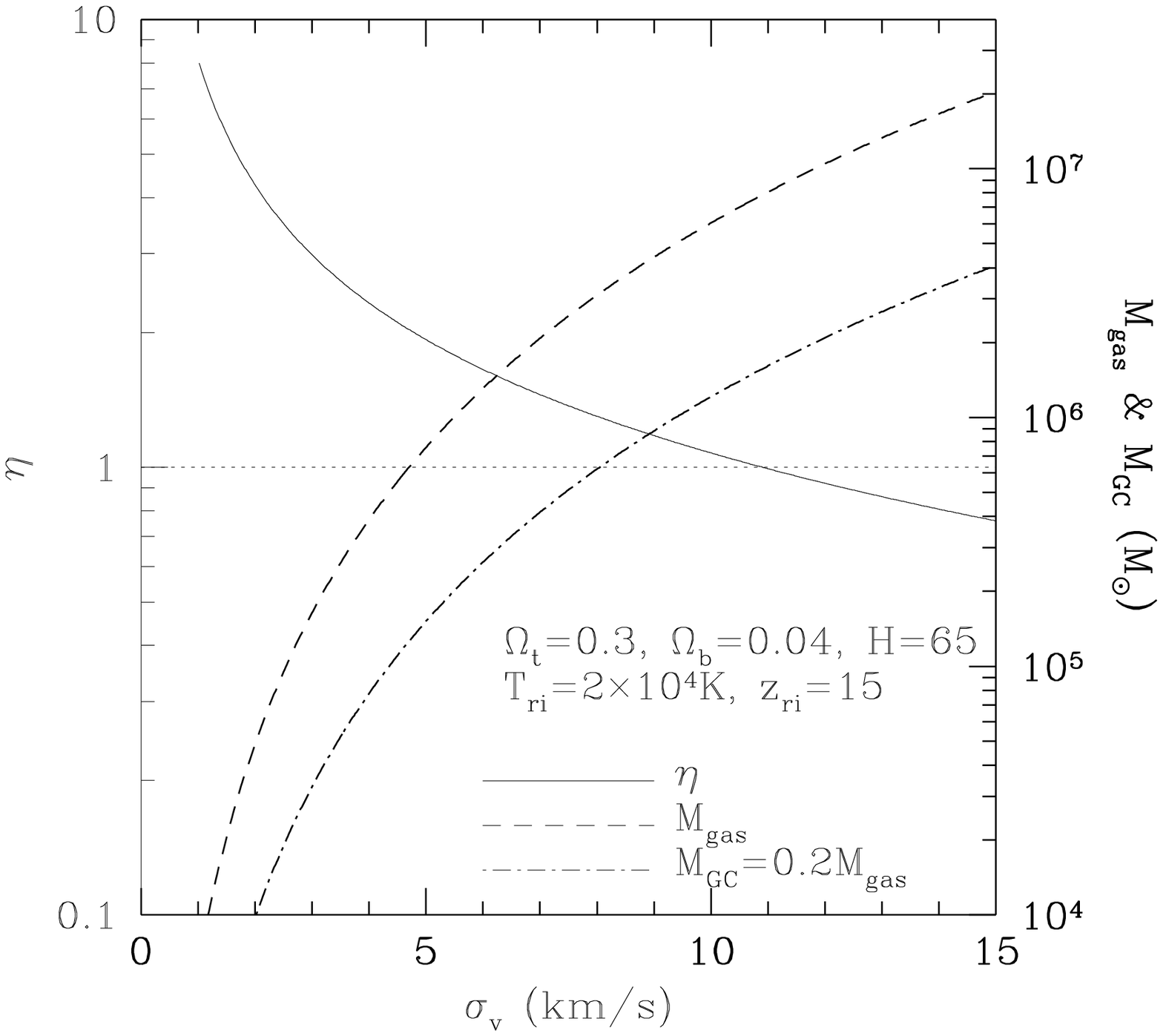}{3.2}{0.45}{-10}{-10} 
\vspace{-0.5cm} 
\figcaption{
shows $\eta$ (equation 8; solid curve) 
as a function of halo velocity dispersion
$\sigma_v$, with a temperature of the photoionized gas of 
$T_{ri}=2\times 10^4~$Kelvin
and $j_{LL,21}=2\times 10^{-3}$ at $z_{ri}=15$.
The long dashed curve is the total collapsed
baryonic mass $M_{gas}$ (see the right vertical axis)
and the dot-dashed curve is $M_{GC}=0.2M_{gas}$,
where $0.2$ is an adopted star formation efficiency.
\label{fig:tau}}
\vspace{\baselineskip}

Figure 1 shows $\eta$ (equation 8) as a function of $\sigma_v$ for
$T_{ri}=2\times 10^4~$Kelvin 
and $j_{LL,21}=2\times 10^{-3}$ at $z_{ri}=15$. 
The adopted value of photoionized gas temperature 
is conservative [Miralda-Escud\'e \& Rees (1994) 
note that the post
ionization temperature may be as high as $5\times 10^4$Kelvin].
We see that halos with velocity dispersion $\sigma_v\le 11~$km/sec
could collapse momentum-wise.
The dashed curve is the total collapsed
baryonic mass $M_{gas}$ (see the right vertical axis)
and the dot-dashed curve is $M_{GC}=0.2M_{gas}$,
where $0.2$ is a putative value for 
star formation efficiency (see below for further discussion).
We see that compact, baryonic systems 
with mass $M_{gas} \le 10^{6}\msun$ may form.

\myputfigure{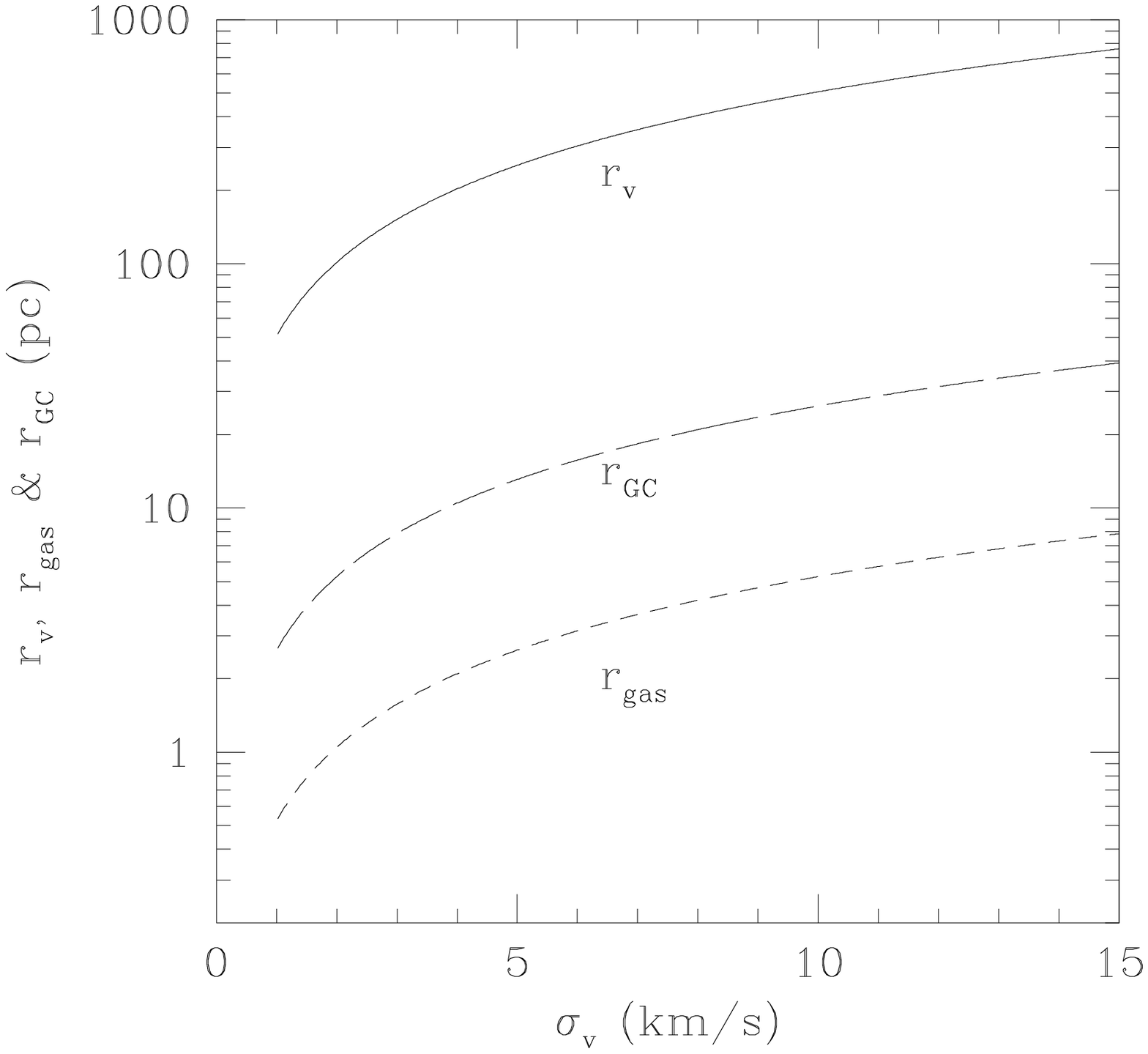}{3.2}{0.45}{-10}{-10} 
\vspace{-0.5cm} 
\figcaption{
shows the initial virial radius (solid curve),
the radius of the compressed gas cloud (short dashed curve)
and the final radius of stellar system (globular cluster)
(long dashed curve), all in proper units,
as a function of velocity dispersion at $z_{ri}=15$.
\label{fig:tau}}
\vspace{\baselineskip}

Let us ask how much compression is required to 
produce globular cluster like systems.
Figure 2 shows the initial virial radius (solid curve)
and the radius of the compressed gas
cloud whose velocity dispersion
is $\sigma_{cm}=\sigma_v/0.2$ (short dashed curve)
as a function of halo velocity dispersion at $z_{ri}=15$.
We see that the gas clouds need to be compressed by 
a factor of $\sim 100$.

A compressed gas cloud will be 
flattened (due to angular momentum) and shocked, 
cool, presumably fragment and form stars.
However, we know little about
high redshift star formation in such systems observationally
or theoretically, 
and simply assume that the star formation efficiency is $20\%$.
From an analysis of Local Group dwarf galaxies
Gnedin (2000a) concludes that star formation at high redshift
is about 4\%, if it was continuous,
but could be significantly higher,
if most of the star formation occurred
well before $10$Gyr ago.
Therefore, our choice of star formation efficiency
seems plausible but uncertainty may be as large as a 
factor of two to five.
Note that the mean baryonic number density in the 
compressed cloud is of order $1000~$cm$^{-3}$,
with the actual final density in a flattened disk probably even higher,
matching that of molecular cloud cores (Friberg \& Hjalmarson 1990;
Goldsmith 1987).
With such a large compression 
the dark matter would become a small fraction (of order $1\%$ for a more
realistic halo density profile) of the final total mass.

The stars formed will likely blow away most
of the remaining gas by supernova explosions
(Dekel \& Silk 1986; Mac Low \& Ferrara 1999).
After the remaining gas is lost as a wind,
the cloud will expand
by approximately a factor of $\sim 1/0.2$ and 
result in a final stellar system with a velocity
dispersion of $\sigma_{GC}=0.2\times \sigma_{cm}=\sigma_v$
and a size shown as the long dashed curve in Figure 2.
We see that the typical size of such a stellar system
is between $2-40~$pc.
Both the assumption that the final stellar system has the same velocity
dispersion as the initial host halo 
and the fudge factor $1/0.2$ in setting $\sigma_{cm}$ 
are rather uncertain,
although these two parameters are degenerate and 
can be condensed into a single parameter.
But small variations would not change the results qualitatively.
Nevertheless, detailed simulations will be necessary to definitively
quantify them.

It is yet unclear whether the required gas compression indicated in Figure 2
is achievable in real halos, even if the momentum
condition $\eta > 1$ (equation 8) is met.
The critical 
issue here is that a finite angular momentum of the halo
gas before compression
(Peebles 1969; White 1984) 
could set a maximum compression factor.
Let us denote the initial spin parameter of the halo (both gas
and dark matter) as 
$\lambda_i\equiv J_i |E_i|^{1/2} G^{-1} M_i^{-5/2}$
(where $J_i$, $E_i$ and $M_i$ are the initial angular momentum,
total energy and total mass, respectively),
then the final spin parameter of the gas cloud is 
(assuming no loss of angular momentum)
$\lambda_f\approx \lambda_i R^{-1} \sigma_{cm}/\sigma_v \approx 40 \lambda_i$
(for the short dashed curve in Figure 2).
Therefore, a typical halo with $\lambda_i\sim 0.05$ 
(Barnes \& Efstathiou 1987; 
Ueda \etal 1994;
Steinmetz \& Bartelmann 1995;
Cole \& Lacey 1996) would
give $\lambda_f\sim 2$.
Clearly, only halos with spin parameter values significant lower
than the typical value
can be adequately compressed to form the globular cluster like systems.
Are there enough number of such low spin halos at $z\sim 15$?

\myputfigure{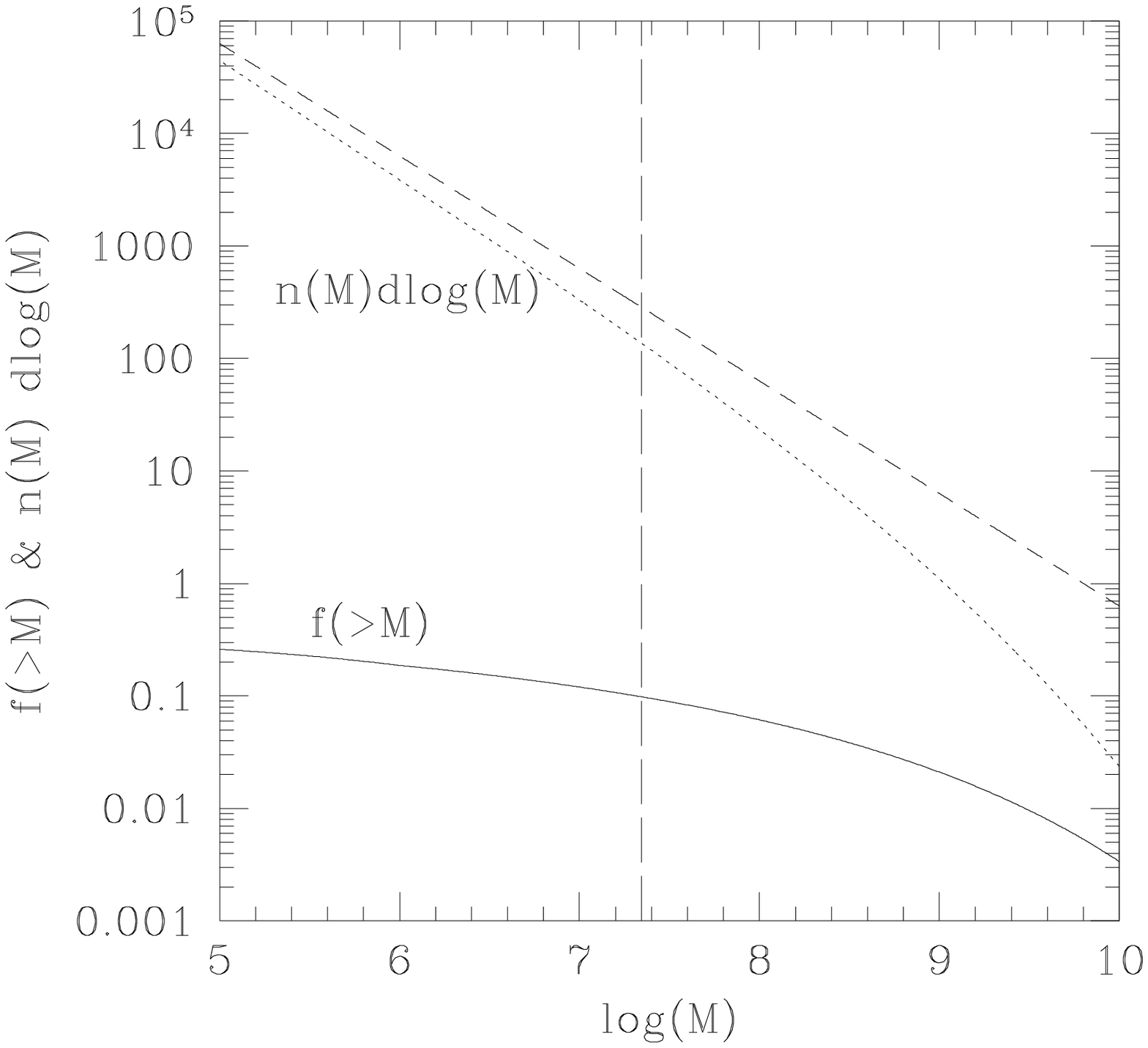}{3.2}{0.45}{-10}{-10} 
\vspace{-0.5cm} 
\figcaption{
shows the mass function of halos at $z=15$ (solid curve) using
the Press-Schechter (1976) formula with
the following cosmological model parameters: 
$\Omega_{CDM}=0.26$, $\Omega_b=0.04$, $\Lambda=0.0$, 
$H_0=65$km/sec/Mpc and $\sigma_8=1.0$.
The vertical line indicates the halo mass with $\eta=1$ as shown in Figure 1. 
The dotted curve is the cumulative mass fraction
in halos, $f(>M)$,
and the dashed curve indicates the
slope of a power-law mass function, $n(M)dM\propto M^{-2}dM$.
\label{fig:tau}}
\vspace{\baselineskip}

Figure 3 shows the cumulative fraction of total mass in halos
at $z=15$ (solid curve) using
the Press-Schechter (1976) formula.
The region left of the vertical (long dashed) line has $\eta\ge 1$ 
(equation 8). 
We see that the (total) mass fraction of halos that 
meet the condition indicated by equation (8) is approximately $15\%$.
As numerical simulations 
have shown that the distribution of the $\lambda_i$
is quite broad with the 80\% range being $\sim 0.01-0.10$
(Ueda \etal 1994;
Steinmetz \& Bartelmann 1995;
Cole \& Lacey 1996);
there are 10\% of halos having $\lambda_i\le 0.01$, which
would give $\lambda_f\le 0.4$ in this model
[when being compressed by a factor of $\sim 100$ to $r_{cm}$
(short dashed curve) in Figure 2].
If $10\%$ of the halos form globular clusters,
the resultant density parameter is
$\Omega_{GC,comp}\approx 0.2\times 1.7 \times \Omega_b \times 15\% \times 10\%
\approx 2\times 10^{-4}$
(where the five terms from left to right are
the star formation efficiency,
the ratio of the total swept-up gas mass 
over the gas mass within the virial radius,
the baryon density,
the mass fraction of halos in question
and the fraction of low spin halos that collapse).
The observed ratio of globular cluster mass over total Galactic baryonic
mass today is approximately $0.25\%$ (Harris \& Racine 1979),
giving $\Omega_{GC,obs}=0.0025 \Omega_*=1.5\times 10^{-5}$,
where $\Omega_*=0.006$ is the stellar mass density 
(e.g., Gnedin \& Ostriker 1992).
Evidently, there are enough low spin halos in the indicated
mass range that could account for the observed halo globular clusters.
Since the slope of the predicted mass function is also in agreement
with observations, one can infer that the expected
number density of globular clusters is consistent with observations.
The issue of exactly what 
the threshold of the spin parameter is in order to
form a globular cluster system can only be settled by detailed
simulations.
Complications such as shell instability may play a role in this regard.
It may be that the initial $\Omega_{GC}$
is significantly larger than the presently
observed value, 
because a large fraction 
(perhaps relatively faster spinning)
of the initial globular clusters have largely 
evaporated or disrupted in time and now are a part 
of the halo stellar population.

The assumption of a thin shell needs further examination.
With the usual shock jump conditions
and equation (7),
we find that at the start of the shock propagation
the shell velocity settles to 
$v_{s}=\sqrt{k T_{ri}/m_p}$ 
and the shock front travels ahead of the shell
with a velocity of 
of $v_{shk} = 4v_{s}/3 $.
The postshock gas temperature would be $T_s=1/3 T_{ri}=6667$Kelvin
and the postshock pressure would be $4p_{ext}/3$.
Therefore, while the shell is bounded at the inner side
by the ram pressure,
the outer boundary will slightly expand initially.
In order to keep the shell thin
the shell gas needs to be cooled.
We compute the gas cooling including all relevant
processes (Haiman \etal 1996) with the assumption 
that the gas has been subject to the radiation 
for a period of $t_{Hubb}$ (the Hubble time at $z_{ri}$)
and the density is $4\rho_v$.
For the gas in question the primary cooling process
is molecular hydrogen cooling and
the primary heating process is photo-heating of hydrogen atoms.
Figure 4 shows the ratio of the cooling time to 
the time scale of the system in question $t_{sys}$
as a function of optical depth at the Lyman limit,
where $t_{sys}\equiv 1\hbox{kpc}/10\hbox{km/sec}\approx 1\times 10^8~$yr.
The incident radiation field 
has $j_{LL,21}=2\times 10^{-3}$ with an index $-2$ 
and the postshock gas 
temperature is assumed to be $5\times 10^3$Kelvin.
We see that gas with a moderate optical depth (shielding) 
of $\tau\sim 4-10$
can cool efficiently (gas at low end of the optical depth
is heated up by photo-heating, while gas at the high end
of the optical depth can not cool efficiently
due to a low abundance of molecular hydrogen).
For a shell of neutral gas we can relate
its mass to the optical depth
as $M(\tau)=3.4\times 10^4 \tau (r/1\hbox{kpc})^2\msun$.
Apparently, adequate shielding may be achieved for the gas 
and gas may be able to cool efficiently.
Note that the gas cooling time remains approximately constant,
assuming it cools isobarically, until it reaches
a few hundred Kelvin.
In addition, self-gravity of the shell will provide
some confinement to keep the shell thin.
More definitive answers can not be given 
without a detailed radiation-hydrodynamic simulation.

\myputfigure{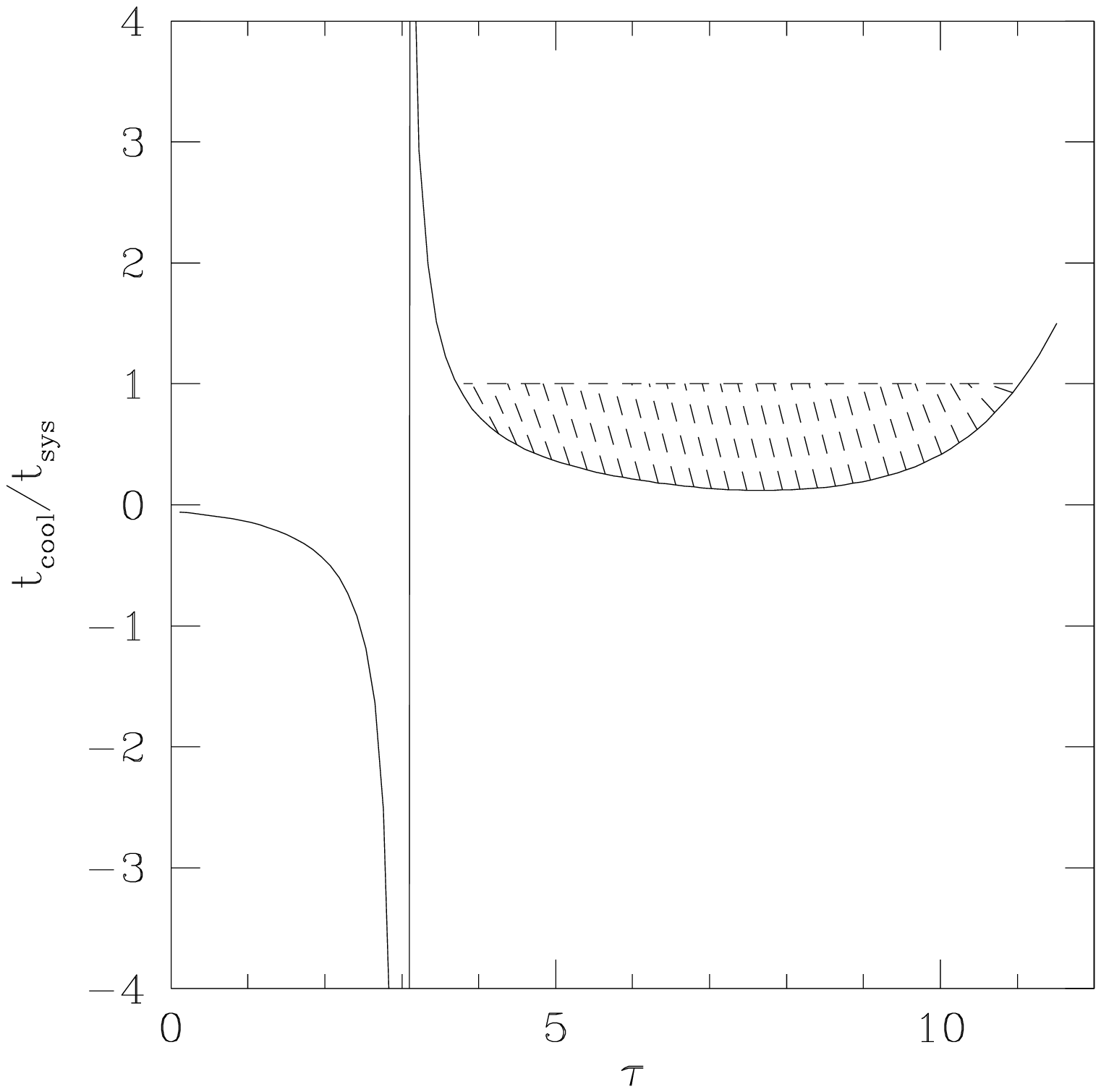}{3.2}{0.45}{-10}{-10} 
\vspace{-0.5cm} 
\figcaption{
shows the ratio of the cooling time to 
the time scale of the system in question
as a function of optical depth at the Lyman limit.
The incident radiation field 
has $j_{LL,21}=2\times 10^{-3}$ with an index $-2$.
A negative ratio means net heating.
The hatched region indicates the range in optical
depth where cooling is efficient.
\label{fig:cool}}
\vspace{\baselineskip}

Finally, we give a dimensional analysis for the mass threshold
of compressed gas clouds.
We use the following line of reasoning. 
The upper mass threshold of the cloud
should be related to the temperature of the external 
ionized gas by
$kT_{ri} = GM_v m_p/r_v$,
where the virial
radius $r_v$ is related to
$M_v$ by $\delta {4\pi\over 3} r_v^3 (1+z_{ri})\Omega_t\rho_{c}(0)=M_v$
($\delta \sim 200$ is the virial overdensity).
The temperature of the external
ionized gas should be related
to hydrogen ionization potential
$kT_{ri}={1\over 2} m_e c^2 \alpha^2$,
where $m_e$ is electron mass, $c$ is the speed of light
and $\alpha$ is the fine structure constant.
Combining these relations and the Friedmman equation we
obtain
$M_{gas} = {R\over 2^{5/2} \pi}{c^3\alpha^3\over H G} ({m_e\over m_p})^{3/2}
= 3\times 10^8 R ({1+z_{ri}\over 15})^{-3/2} \Omega_t^{-1/2}h^{-1}\msun$,
where $H$ is the Hubble constant at $z_{ri}$ and $R$ is the global
gas to total mass ratio.
It is seen that the upper mass threshold is around 
$\sim 10^{7.5}\msun$, 
consistent with the detailed calculations given above.

\section{Characteristics of the Globular Clusters}

The dotted curve in Figure 3 is the computed mass function
of halos, $n(M)$ (in units of $h^{3}Mpc^{-3}$),
using the Press-Schechter formula and 
the dashed curve a power-law mass function, $n(M)dM\propto M^{-2}dM$.
A globular cluster mass function of 
$n(M) dM = M^{-2}dM$ seems to be borne out naturally in this picture,
in excellent agreement with observations 
for globular clusters more massive than $10^5\msun$
(Surdin 1979; Racine 1980; Richtler 1992).
The observed shallower slope at lower mass may be
due to disruptive effects 
during subsequent dynamical evolution of globular clusters,
which tend to work against low mass systems
(Chernoff, Kochanek, \& Shapiro 1986; 
Chernoff \& Shapiro 1987;
Spitzer 1987; 
Aguilar, Hut, \& Ostriker 1988;
Chernoff \& Weinberg 1990;
Harris 1991;
Goodman 1993; Gnedin \& Ostriker 1997).
We note, however, there may be a cutoff in the 
velocity dispersion of globular clusters at $\sigma_{GC}\sim 2$km/s,
corresponding to $\sigma_{cm}\sim 10$km/s, below which
cooling and thus star formation are less efficient 
at the absence of atomic cooling.

It is likely that a significant 
fraction of the initial globular clusters
may possess a significant amount of angular momentum
and their initial shapes may be significantly flattened.
While much more work is needed to quantify the dynamical evolution
of rotating many-body stellar systems (e.g., Einsel \& Spurzem 1999),
Akiyama \& Sugimoto (1989) show that initially rotating 
stellar systems tend to become rounder with time.
Although it is not possible at present to make
useful comparisons with observations,
some initial rotation of the globular clusters
might be preferred by extant observations (e.g., White \& Shawl 1987),

Simulations of Ueda \etal (1994) (also Cole \& Lacey 1996)
show a correlation between average spin parameter $\bar\lambda_i$ and 
halo mass, $\bar\lambda_i \propto M_v^{-0.2\pm 0.1}$,
for massive (galaxy size) systems that they simulate.
If we assume that for the halos in question
a similar correlation exists, such as $\bar\lambda_i \propto M_{v}^{-1/6}$,
and that gas clouds of all halos are compressed 
to a state of comparable final spin parameter,
then the average final cloud radius 
$\bar r_{GC}\propto r_v \bar\lambda_i^2 \propto r_v M_{v}^{-1/3} \propto M_v^0$ 
(because $r_v\propto M_{v}^{1/3}$),
i.e., the average final radius of the globular clusters
is independent of the initial mass of the halo,
in agreement with the observed lack of correlation between
the half-light radius and the luminosity (Djorgovski \& Meylan 1994).
Of course, since there is a range in the initial spin parameter values,
there will be a corresponding spread in the sizes of globular clusters,
as observed.
Since the final velocity dispersion of a
globular cluster is $\sigma_{GC}\propto (M_v/\bar r_h)^{1/2}$,
the independence of $\bar r_h$ of $M_v$ dictates
$\sigma_{GC}\propto M_v^{1/2} \propto L^{1/2}$, in agreement with
the observed correlation $\sigma_{GC}\propto L^{0.6\pm 0.15}$
(as well as the observed correlation with 
surface brightness
$\sigma_{GC}\propto I_h^{0.45\pm 0.05}$) (Djorgovski \& Meylan 1994).
A prediction that may be made
is that the logarithmic dispersion in $\sigma_{GC}$ in the 
$\sigma_{GC}-L$ correlation should be about half the size
of the logarithmic dispersion in the size of globular clusters,
which seems to be in agreement with observations (Djorgovski \& Meylan 1994).

There appears to exist a tendency for halos with 
relatively large $\lambda$ to reside
preferentially in low density regions in simulations
(e.g., Ueda \etal 1994).
This effect would conceivably produce the observed
correlation of the specific frequency of globular clusters
with galaxy type in that early type galaxies
have a high number of globular clusters per unit galactic luminosity
than late type galaxies (Harris 1991).
This arises because
early type galaxies originate from initial higher density 
peaks and would have (on average)  
lower spins and thus higher specific frequencies.
Quite intriguing is that this trend might have some bearing
on the formation of spiral and elliptical galaxies and Hubble sequence.
But within a similar galaxy type, 
the number of globular clusters
in a galaxy may simply be proportional to the mass or luminosity of the galaxy,
as observed (Hanes 1977; Harris \& Racine 1979),
since the mass of a galaxy basically indicates 
the size of the initial (comoving) region from which it has
collected matter hence proportionally the number of globular clusters.
More certain is that the range of properties of globular clusters
such as size, luminosity and velocity dispersion
should be quite uniform across
all galaxies, independent of size, type, age, etc,
as observed (Harris 1991). 
This is due to the fact 
that the formation of globular clusters in the present model
has little correlation with later large-scale density fluctuations that
form galaxies (except the correlation between spin and overdensity mentioned
above).

Chemical enrichment by Pop III stars, which is computed
to be $\sim 10^{-3.5}\zsun$ (Ostriker \& Gnedin 1996),
is much lower than the observed stellar metallicity of 
metal poor halo globular clusters of $\le 10^{-1.2}\zsun$ (Zinn 1985).
It thus appears that some low level of self-enrichment
has to occur in order to account for the observed metallicity of globular
clusters.
While a detailed treatment of star formation in 
the proposed globular clusters is not possible,
the two-generation star formation scenario 
(Cayrel 1986; Brown \etal 1991,1995;
Zhang \& Ma 1993; Parmentier \etal 1999)
could work in this picture.
We note that the escape velocity of a compressed cloud
(where star formation occurs)
is larger than its corresponding globular cluster by a factor of $5$
(this is rather uncertain, though).
Therefore, confinement of the first generation of supernovae
(Dopita \& Smith 1986)
may be more readily accommodated than in other scenarios.
Then, the second generation of stars formed in a massive starburst
(with the assumed formation efficiency of 20\%)
completely blows away the remaining gas,
except for perhaps the most massive globular clusters
(Dekel \& Silk 1986; Mac Low \& Ferrara 1999),
for which further self-enrichment may take place.
This is in accord with the observed internal
chemical homogeneity of most globular clusters (cf. Larson 1988),
with the exception of the most massive
systems such as $\omega$ Cen where some chemical inhomogeneities
are observed (Cohen 1981).

While halo globular clusters depicted 
here should be, on average,
metal poor, as observed (Zinn 1985),
complex star formation histories quite likely
introduce a large dispersion in metallicity among globular clusters.
Thus it may be expected that larger galaxies 
that formed by collecting matter in a larger (comoving) region
are expected to display a larger variance in metallicity, in agreement with
observations (Harris 1991).
Since the metallicity distribution is asymmetric
and bounded by $Z=0$ at the low end,
a large variance gives a larger mean, 
interestingly consistent with observations (Harris 1991).
On the other hand, since metallicity does not
play a noticeable role in the formation of 
individual globular clusters (i.e., dynamic collapse of the gas clouds)
in this model,
it should be expected to show little correlation with any other quantities
of globular clusters,
such as luminosity, size, velocity dispersion,
in agreement with observations (Djorgovski \& Meylan 1994).

If the universe went through a second, separate 
reionization phase of singly ionized helium at 
lower redshift ($z\sim 3$), as recent observations hint
(e.g., Reimers \etal 1997),
a similar radiation I-front will sweep through halos of neutral gas.
However, due to much lower density and lower abundance of 
helium (compared to hydrogen) at lower redshift,
the speed of the I-front is likely to be larger
than the shock speed until well within the virial radius;
i.e, the I-front will remain as an R-type and no significant compression
will occur.
Therefore, our model indicates that the age spread of metal poor
globular clusters should be less than $1~$Gyr,
consistent with recent observations (Rosenberg \etal 1999).

Finally, it seems inevitable that a fraction of 
the original globular clusters may be in
the intergalactic space where no galaxies were formed
to collect them.
These intergalactic globular clusters
are possibly at a somewhat younger
dynamic state than Galactic halo globular clusters
due to lack of external dynamical effects.
The exact number density of intergalactic globular clusters
depends sensitively on the angular momentum distribution
of halos in low density regions and requires
more detailed calculations for quantification.
No doubt it will be important to search for them with
the next generation of large telescopes.

\section{Conclusions}

Radiation from stars in galaxies with mass $M_t>10^9\msun$ or quasars 
are thought to be mostly likely responsible for reionizing 
the universe at redshift $z\sim 6-20$.
However, a large fraction of mass
resides in density concentrations on smaller mass scales, i.e.,
gas rich sub-galactic halos with mass $M_t<10^8\msun$.
These smaller systems can not cool to form stars without 
an external trigger.
Reionization of the universe has 
an interesting dynamic effect on these halos.
The rise of the external radiation field causes a synchronous 
propagation of an inward, convergent shock towards each halo.
We show that, for halos with a velocity dispersion $\sigma_v\le 11~$km/s,
the resident gas of mass $M_b\sim 10^4-10^{7}\msun$ will be
compressed.
For a significant fraction of these halos that have 
relatively low initial angular momentum ($\lambda \le 0.01$),
a compression of a factor of $\sim 100$ in radius is possible
and compact
self-gravitating baryonic systems would form.
Such gaseous systems fragment to form stars and
become stellar systems of size $\sim 2-40~$pc,
velocity dispersion $\sim 1-10$km/s and a
total stellar mass of $M_s\sim 10^3-10^{6}\msun$.

Their expected properties all seem to be in agreement with the observed 
salient features of halo globular clusters, as
discussed in the previous section.
Angular momentum distribution of small halos at high redshift
plays a critical role in determining many of the properties
of globular clusters.
The expected number density of halo globular clusters
is consistent with what is observed.
The mass function of slope $\sim -2$ at the high mass end
is predicted by the model.
Strong correlation between velocity dispersion and luminosity 
(or surface brightness) and lack of correlation between velocity
dispersion and size are expected.
Globular clusters should display similar properties, regardless
of the host galaxy type, age, size or luminosity.
Metallicity is, on average, expected to be low
and should not correlate with any other quantities of globular clusters,
except that a larger dispersion of metallicity among globular clusters
is expected for larger galaxies.
A trend of specific frequency with galaxy type
may be produced in the model.
We propose that these stellar systems 
represent the initial state of the 
presently observed halo globular clusters.

It will be important to systematically search for high redshift 
as well as intergalactic globular clusters, because
they will provide tests of the proposed model
and could potentially shed light on the theory of universe
reionization and the general formation picture of galaxies.
Also important is to make detailed radiation-hydrodynamic
simulations with high resolution and superior shock
resolving power to study the reionization process and
follow the dynamics of gas collapse.
Both are formidable challenges.

\acknowledgments
I thank Bruce Draine, Jerry Ostriker and Michael Strauss 
for a critical reading of the manuscript and many cogent comments,
Stu Shapiro for suggesting a dimensional analysis,
Zoltan Haiman for allowing me to use his cooling program,
and Rennan Barkana and Jordi Miralda-Escud\'e for useful discussion.
This research is supported in part by grants AST93-18185
and ASC97-40300.

\end{document}